\documentclass{ws-ijmpd}
\usepackage[super,compress]{cite}
\usepackage{color}
\usepackage{hyperref}

\newcommand{\bq}{\begin{equation}}
\newcommand{\eq}{\end{equation}}
\newcommand{\bqa}{\begin{eqnarray}}
\newcommand{\eqa}{\end{eqnarray}}
\def\ba#1\ea{\begin{align}#1\end{align}}

\def\lsim{\, \lower .75ex \hbox{$\sim$} \llap{\raise .27ex \hbox{$<$}} \,}

\begin{document}

\markboth{Zhenjie Liu, Haitao Miao}
{}

%%%%%%%%%%%%%%%%%%%%% Publisher's Area please ignore %%%%%%%%%%%%%%%
%
\catchline{}{}{}{}{}
%
%%%%%%%%%%%%%%%%%%%%%%%%%%%%%%%%%%%%%%%%%%%%%%%%%%%%%%%%%%%%%%%%%%%%

\title{Update constraints on neutrino mass and mass hierarchy in light of dark energy models}

\author{Zhenjie Liu$^2$}

%\address{School of Physics and Astronomy, Sun Yat-sen University, \\
 %Zhuhai, China\\
\address{liuzhj26@mail2.sysu.edu.cn}

\author{Haitao Miao$^{1, 2}$}

\address{miaoht3@mail2.sysu.edu.cn}

\address{
1. Key Laboratory of Space Astronomy and Technology, National Astronomical Observatories,Chinese Academy of Sciences, \\
Beijing, China\\
2. School of Physics and Astronomy, Sun Yat-sen University, \\
Zhuhai, People's Republic of China\\}
%miaoht3@mail2.sysu.edu.cn}

\maketitle

%\begin{history}
%\received{18 May 2020}
%\revised{11 July 2020}
%\accepted{17 July 2020}
%\published{7 September 2020}
%\end{history}

\begin{abstract}
Combining cosmic microwave background (CMB) data from Planck satellite data, Baryon Acoustic Oscillations (BAO) measurements and Type Ia supernovae (SNe Ia) data, we obtain the bounds on total neutrino masses $M_\nu$ with the approximation of degenerate neutrino masses and for three dark energy models: the cosmological constant ($\Lambda$CDM) model, a phenomenological emergent dark energy (PEDE) model and a model-independent quintessential parameterization (HBK). The bounds on the sum of neutrino masses $M_\nu$ depend on the dark energy (DE) models. In the HBK model, we confirm the conclusion from some previous work that the quintessence prior of dark energy tends to tighten the cosmological constraint on $M_\nu$. On the other hand, the PEDE model leads to larger $M_\nu$ and a nonzero lower bound. Besides, we also explore the correlation between three different neutrino hierarchies and dark energy models.
  
\end{abstract}


\ccode{PACS numbers:95.35.+d}
%\tableofcontents

\section{Introduction}\label{intro}
The standard model of particle physics predicts that neutrinos are massless, whereas the discovery of neutrino oscillations, a phenomenon that neutrinos can switch their flavour to others, suggests that they are massive. The neutrino oscillation experiments can only accurately measure the squared mass differences between two types of individual neutrino instead of their absolute masses. From neutrino oscillation data, we know the values of mass-squared splittings: $\Delta m_{21}^2 \equiv m_2^2-m_1^2 \approx 7.54_{-0.22}^{+0.26} \times10^{-5}{\rm eV}^2$, $|\Delta m_{31}^2| \equiv |m_3^2-m_1^2| \approx 2.46_{-0.06}^{+0.06} \times10^{-3}{\rm eV}^2$\cite{Tanabashi2018}. As we do not know whether $\Delta m_{31}^2$ is positive or negative, two kinds of neutrino mass ordering, normal hierarchy (NH, $m_3 \gg m_2 \textgreater m_1$) and inverted hierarchy (IH, $m_2 \textgreater m_1 \gg m_3$), are possible. The degenerate hierarchy (DH, $m_1 = m_2 = m_3$) is also widely used in the cosmological parameter estimations, while it is not physical.  Furthermore, different orderings of neutrino mass have different minimums of their total masses ($M_\nu=\sum_i m_i$): for NH scheme given by $\left(M_\nu\right)_{min} = \sqrt{\Delta m_{21}^2} + \sqrt{\Delta m_{31}^2}  \approx 0.06{\rm eV}$, while for IH scheme we have $\left(M_\nu \right)_{min} = \sqrt{|\Delta m_{31}^2|} + \sqrt{ |\Delta m_{31}^2| - \Delta m_{21}^2} \approx 0.1{\rm eV}$. Hence, which mass  hierarchy is more favoured by the nature is important for further understanding the evolution of the universe. Nowadays, cosmology has been an effective tool to detect neutrinos, which can provide the most robust bounds on the neutrino masses. The tight bounds on the sum of neutrino masses may provide a solution to the neutrino mass hierarchy problem. 

Massive neutrinos play an important role in some cosmic phenomena, such as the formation of the large-scale structure (LSS), the big bang nucleosynthesis (BBN), the anisotropy of the cosmic microwave background (CMB) etc~\cite{2002PhR...370..333D,2006PhR...429..307L}. Neutrinos have distinct effects on the evolution of the universe and leave traces on the CMB power spectrum and LSS power spectrum, so we can find their signatures in cosmological observations. Gerstein and Zeldovich were the first to derive the cosmological upper limit on the total neutrino masses~\cite{2002PhR...370..333D}. Since then, plenty of further investigations have been carried out in the literature \cite{2002PhR...370..333D,2006PhR...429..307L,2017PhRvD..96l3503V,2012arXiv1212.6154L,2016PhRvD..93c3001G,2017JCAP...02..052A,2016PhRvD..94h3522G,2018JCAP...03..011G,2018PhRvD..98l3526G,2018JCAP...09..001V}. Up to now, current cosmological observations are primarily sensitive to the sum of neutrino masses. However, the bounds on the sum of neutrino masses depend on dark energy models~\cite{2018JCAP...09..017C,2019arXiv190712598C}. Dark energy powered the accelerating expansion of the universe in late times. Although current observational data tend to the $\Lambda$CDM model, we still could not rule out the quintessence and phantom model of dark energy. Different dark energy models will put different bounds on the sum of neutrino masses. According to the recent works\cite{2017SCPMA..60f0431Z,2018PhRvD..98h3501V,2016PhLB..752...66C,2018arXiv180702860C}, the constraints on $M_\nu$ in quintessence model are found to be tighter than those obtained in $\Lambda$CDM model. There are also studies that talk about neutrino hierarchy from cosmology~\cite{2017PhRvD..95j3522Y,2018PhRvD..97d3510L,2017PhLB..775..239G,2019arXiv190700179Z,2017PhRvD..96l3503V,2017arXiv170304585S,2016PhRvD..94h3519W,2017SCPMA..60f0431Z,2017JCAP...06..029S,2018JCAP...03..011G,2018JCAP...04..047H,2018FrASS...5...36M,2017PhRvD..95d3512G}. In most studies, the current cosmological data are not sensitive enough to distinguish the normal hierarchy and the inverted hierarchy, however there is a slight preference to NH~\cite{2017PhLB..775..239G,2018FrASS...5...36M,2018JCAP...03..011G,2019arXiv190700179Z,2017PhRvD..96l3503V,2016PhRvD..94h3519W,2019arXiv190712598C,2017SCPMA..60f0431Z,Mahony2020}.

Recently, a simple phenomenological emergent dark energy (PEDE) model has been proposed to resolve the Hubble Tension \cite{2019ApJ...883L...3L,2019arXiv190712551P}. This model assumes that dark energy does not exit effectively in the past but emerges at later time. The equation of state (EOS) $w_{\rm de}$ of dark energy goes from $-\frac{2}{\rm 3 ln 10}-1$ in the past to $-1$ in the future. Additionally, an analytic approximation of EOS has been derived based on a minimally coupled and slowly or moderately rolling quintessence field $\phi$ with a smooth potential $V(\phi)$ (HBK model)~\cite{2011ApJ...726...64H}.  Based on the PEDE model, the HBK model and the $\Lambda$CDM model, and combined with CMB data from Planck 2018, BAO measurements and SNe Ia, we investigate the bounds on the sum of neutrino masses $M_\nu$ with the approximation of degenerate neutrino masses. We also explore the correlation between three different neutrino hierarchies and dark energy models.

This article is organized as follows. Sec.\ref{two} describes the methodology and the observational data used in our analysis. We discuss our results in Sec.\ref{three}. In Sec.\ref{four}, conclusion and discussion are given.

\section{Methodology and Datasets}\label{two}
In our analysis, we apply three dark energy models with different types of dark energy evolution: $\Lambda$CDM, PEDE, HBK. To perform bayesian analysis of the cosmological dataset, we modify the publicly available markov chain monte carlo (MCMC) package CosmoMC \cite{2002PhRvD..66j3511L} with the Boltzmann solver CAMB~\cite{2000ApJ...538..473L}. We study the impacts of neutrinos on the CMB temperature spectrum and the matter power spectrum. The priors on the main cosmological parameters used for all models are listed in Tab.~\ref{priortab}. Here is an introduction to the dark energy models and datasets.

\begin{table}[ht]
%\begin{center}
% \renewcommand{\arraystretch}{1.4}
\tbl{Priors on the main cosmological parameters included in this paper.}
{\begin{tabular}{c c}
\hline
Parameters & Prior\\
\hline
$\Omega_bh^2$  &[0.005, 0.1] \\
$\Omega_ch^2$   &[0.01, 0.99 \\
$\Theta_s$    &[0.5, 10] \\
$\tau$   &[0.1, 0.8] \\
$n_s$  & [0.8, 1.2] \\
${\rm ln}(10^{10}A_s)$  &[1.6, 3.9] \\
$M_\nu$ (eV) & [0, 3]\\
$\epsilon_s$  &[0, 0.5] \\
$\epsilon_{\phi\infty}$  &[0, 1] \\
$\zeta_s$     & [-1, 1]\\
$H_0$ (km/s/Mpc)& [40, 100]\\
\hline
\end{tabular}}
%\end{center}
\label{priortab}
\end{table}

\subsection{Dark Energy Models}
\textbf{The $\Lambda$CDM + $M_\nu$ model.} The parameter space of the $\Lambda$CDM model is
\begin{equation}\mathcal{P}\equiv\{\Omega_bh^2,\Omega_ch^2,100\Theta_{\rm MC},\tau,n_s,{\rm ln}(10^{10}A_s),M_\nu\},
\end{equation}
which has the minimum number of parameters compared to other models.  $\Omega_bh^2$ and $\Omega_ch^2$ describes the baryon and cold dark matter densities today. $\Theta_{MC}$ is an approximation to the angular size of sound horizon at the time of decoupling. $\tau$ is the optical depth due to re-ionization. $n_s$ and $A_s$ refer to the spectral index and the amplitude of initial power spectrum related to early universe cosmology. The EOS for $\Lambda$CDM model remains a constant, i.e. $w_{\rm de} = -1$.

\textbf{The PEDE model.} It is a phenomenological model of emergent dark energy, which did not exist effectively in the past until later time~\cite{2019ApJ...883L...3L,2019arXiv190712551P}. It has the same parameter space $\mathcal{P}$ as the minimal $\Lambda$CDM cosmology, namely it does not use any additional degrees of freedom. The EOS given by
\begin{equation}\label{eospede}w_{\rm de} = -1 - \frac{1}{3{\rm ln}10} \times \big[1 - {\rm tanh}({\rm log_{10}}a)\big]. \end{equation}
The dark energy evolves as
\begin{equation}\rho_{\rm de} = \rho_{\rm de,0} \times \big[1 + {\rm tanh}({\rm log_{10}}a)\big], \end{equation}
where $a$ is a scale factor normalized to unity today. From Eq.\ref{eospede}, we can derive that the dark energy state equation goes from $-\frac{2}{3{\rm ln}10}-1$ in the past to $-1$ in the future.

\textbf{The HBK model.} We use an analytic approximation of $w_{\rm de}$ in quintessence models proposed by Huang et al.~\cite{2011ApJ...726...64H}. It fits well the ensemble of trajectories for a wide class of potentials $V(\phi)$ with three additional parameters compared with the $\Lambda$CDM model. The EOS takes the following form
\begin{equation}\label{eos}
  \begin{aligned}
    w_{\rm de} = &-1+\frac{2}{3}\bigg\{\sqrt{\epsilon_{\phi\infty}}+(\sqrt{\epsilon_s} - \sqrt{2\epsilon_{\phi\infty}}) \times \bigg[F\bigg(\frac{a}{a_{eq}}\bigg) + \zeta_s F_2\bigg(\frac{a}{a_{eq}}\bigg)\bigg]\bigg\}^2,
  \end{aligned}
\end{equation}
where, the parameter $\epsilon_s$ describes the slope of the potential $V(\phi)$   at $a=a_{eq}$, i.e. the dark energy density is equal to matter density. The tracking parameter $\epsilon_{\phi\infty}$ characterizes the  curvature of the scalar-field logarithm potential at the pivot, and the running parameter $\zeta_s$ is the initial velocity of the scalar field.  The functions F and $F_2$ are given by
\begin{equation}F(x) \equiv \frac{\sqrt{1+x^3}}{x^{3/2}} - \frac{{\rm ln}[x^{3/2} + \sqrt{1+x^3}]}{x^3},
\end{equation}

\begin{equation}F_2(x) \equiv \sqrt{2}\bigg[1 - \frac{{\rm ln}(1+x^3)}{x^3}\bigg] - F(x),
\end{equation}
respectively. The detailed derivation can be seen in
Refs. \cite{2011ApJ...726...64H, 2018ApJ...868...20M}.

\begin{figure*}[ht]
    \centering
    \includegraphics[width=0.5\textwidth]{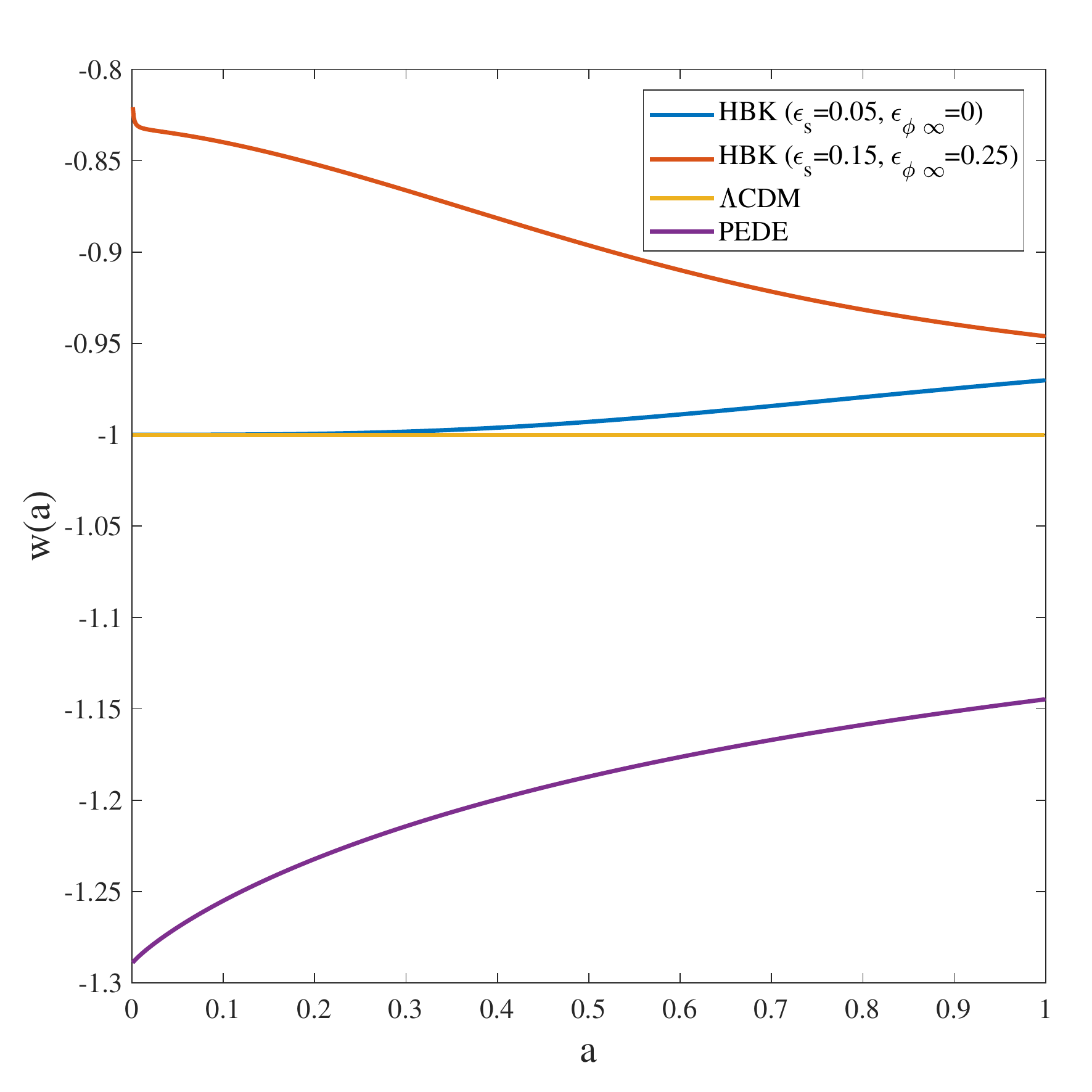}
    \caption{The time-dependence of the dark energy equation of state in the three models.}
    \label{w_de}
\end{figure*}

We show in Fig. \ref{w_de} the time-dependence of the dark energy equation of state in the three models. For HBK model, we take $\epsilon_s=0.05$, $\epsilon_{\phi \infty}=0$, $\zeta_s=0$ and $\epsilon_s=0.15$, $\epsilon_{\phi \infty}=0.25$, $\zeta_s=0$ as examples, which respectively correspond to thawing and tracking models\cite{2011ApJ...726...64H}.

\subsection{Dataset}

We process the most recent datasets from Planck 2018~\cite{2019arXiv190712875P} in combination with other low-redshift observations. We use the Planck 2018 CMB low-l ($2 \leq l \leq 29$) and high-l ($30 \leq l \leq 2508$) TT likelihood, high-l E mode polarization and temperature-polarisation cross correlation likelihood, and low-l E mode polarization likehood. We also include the CMB lensing data. The low-redshift observations contain the Baryon acoustic oscillations (BAO) measurements and Type Ia supernovae (SNe Ia) data. BAO measurements cover 6dFGS~\cite{2011MNRAS.416.3017B}, SDSS-MGS~\cite{2015MNRAS.449..835R} and BOSS DR12~\cite{2017MNRAS.470.2617A} surveys. SNe Ia data are taken from the latest Pantheon Sample~\cite{2018ApJ...859..101S}, including the information of 1048 type Ia supernovae in the range of redshift (0.01 $\textless$ $z$ $\textless$ 2.3).

\section{results}\label{three}
In this section, we present the bounds on neutrino masses for three models and we analyse the correlation between three different neutrino hierarchies and dark energy models. We use three data combinations mentioned above, CMB + BAO + SNe Ia, in all the models.

\begin{figure*}[!htb]
    \centering
    \includegraphics[width=1\textwidth]{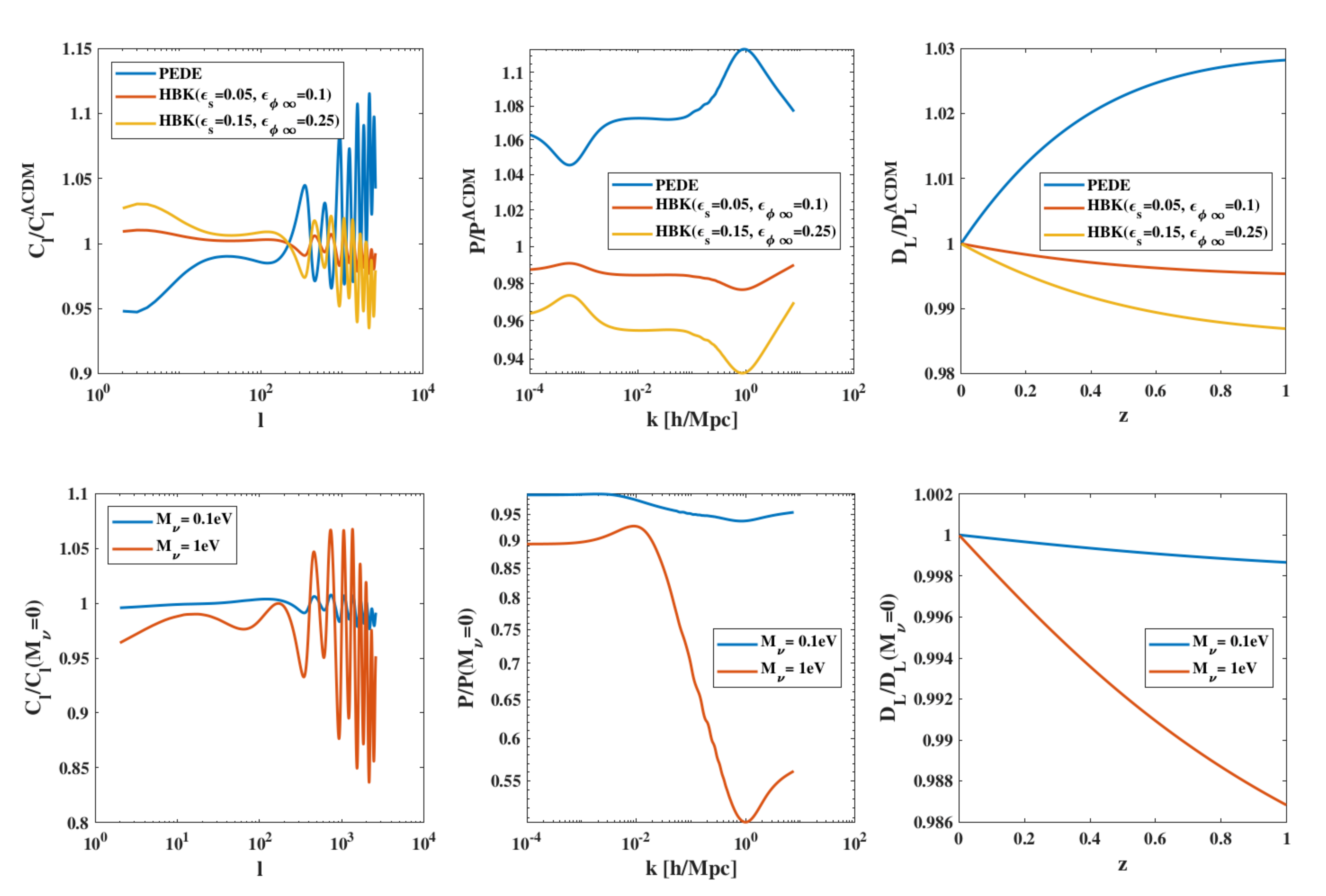}
    \caption{The upper panels show the ratios of different kinds of dark energy EOS to $\Lambda$CDM for the CMB temperature power spectrum $C_l^{TT}$, the matter power spectrum and the luminosity distance. The impacts of neutrino mass showed in the lower panels. They are obtained by keeping fixed all cosmological parameters but those explicitly mentioned in the figure.}
    \label{epsmodels}
\end{figure*}

\subsection{Bounds on neutrino masses}

\renewcommand\tabcolsep{5.0pt}
\begin{table}[ht]
%\begin{center}
\tbl{The 95\% CL bounds on the sum of neutrino masses $M_\nu$ with degenerate approximation in HBK model, $\Lambda$CDM model and PEDE model, respectively.}
{\begin{tabular}{c  c  c  cp{4cm}}
\hline
\hline
&$\Lambda$CDM&HBK & PEDE\\
\hline
$M_\nu$ (eV) &$\textless$ 0.114&$\textless$ 0.087&$0.2123_{-0.1367}^{+0.1293}$\\
\hline
\end{tabular}}
%\end{center}
\label{mnutab}
\end{table}

Here, we talk about the sum of neutrino masses within the assumption of DH. In Tab.\ref{mnutab},  we list the bounds on the sum of the neutrino masses for three dark energy models. When using Planck 2018 data for $\Lambda$CDM model, we find that the upper bound on $M_\nu$ is $0.114\ eV$ at 95\% confidence level (CL), which is more stringent than the result from Vagnozzi et al.~\cite{2018PhRvD..98h3501V} using Planck 2015 data. Planck 2018 improved measurement of $\tau$, it helps break the degeneracy between $\tau$ and $M_\nu$, which leads to tighter bounds on $M_\nu$. Besides, various systematic effects that present in the high-l polarization spectra of Planck 2015 have been corrected by Planck 2018, which also help to obtain a tighter bounds on $M_\nu$.  Moreover, we find a tighter upper bound on $M_\nu < 0.087 \ eV$ at 95\% CL for HBK model, which confirms the conclusion that the constraints on the sum of neutrino masses in quintessence models are tighter than the results in $\Lambda$CDM model. In particular, for PEDE model, we find the constraints on $ M_\nu$ is $0.2123_{-0.1367}^{+0.1293}$ at 2$\sigma$. We note that PEDE model tends to have a larger value of the sum of neutrino masses and it even gives a nonzero lower bound. As is studied in Ref.\cite{2005PhRvL..95v1301H}, there is an anti-correlation between $M_\nu$ and $w_{\rm de}$, a smaller $w_{\rm de}$ tends to a larger $M_\nu$.

In Fig. \ref{epsmodels}, we figure out the theoretical prediction on the CMB temperature spectrum $C_l^{TT}$, the matter power spectrum $P(k)$ and the luminosity distance $D_L$ by showing their relativeness between $\Lambda CDM$ model and other models in the upper panels, to explore the impacts of $w_{\rm de}$. The lower panels show the impacts of neutrinos. For HBK model, we take $\epsilon_s = 0.05, \epsilon_{\phi\infty} = 0.1, \zeta_s = 0$ and $\epsilon_s = 0.15, \epsilon_{\phi\infty} = 0.25, \zeta_s = 0$ with other parameters kept fixed. Notice that, when $\epsilon_s = 0$, $\epsilon_{\phi\infty} = 0$ and $\zeta_s = 0$, the HBK model corresponds to $\Lambda$CDM model. We also display the case of PEDE model for comparison. According to Eq.~\ref{eos}, $w_{\rm de}$ would increase as the increase of $\epsilon_s$. For flat spacetime, the Hubble parameter is
\begin{equation}
 H(z) =H_0\sqrt{\Omega_m(1+z)^3 + \Omega_{\gamma}(1+z)^4 +\Omega_{de}(1+z)^{3(1+w)} + \frac{\rho_\nu(z)}{\rho_{crit,0}}} \ ,
\end{equation}
where  $\Omega_m$, $\Omega_{\gamma}$ and  $\Omega_{de}$ represent density parameters of matter, photons and dark energy, respectively. $\rho_\nu(z)$ represents total neutrinos density and $\rho_{crit,0}$ denotes critical density today. The luminosity distance is
\begin{equation}
D_L = (1+z) \int^z_0\frac{dz'}{H(z')} \ .
\end{equation}
When the EOS of dark energy $w_{de}$ decrease, dark energy contributes to the total energy density budget decrease which also leads to a decrease in $H(z)$. From the upper panels, we can see the significant variations for low-l tail of the CMB temperature, the matter power spectrum $P(k)$ and the luminosity distance as $w_{de}$ changes. We have known from the Refs.~\cite{2012arXiv1212.6154L,2017MNRAS.469.1713Z,2016PhLB..752...66C} that a larger $M_\nu$ will cause the suppression on the CMB temperature spectrum at the low multipoles (late Integrated Sachs-Wolfe effect) and the matter power spectrum. From the lower panels, we can see those variations as neutrino masses change. The anti-correlation between $w_{de}$ and $M_\nu$ can be explained by the compensation to the effects on the acoustic peak scale of $C_l^{TT}$. Fixed acoustic peak, the decreased $H(z)$ by decreasing $w_{de}$ can be compensated by increasing $M_\nu$. Therefore, at fixed cosmological parameters, one would have less energy density in dark energy in the PEDE model. This is the reason why PEDE model favours larger neutrino masses. This also explains the result that the tighter constraints on $M_\nu$ in HBK model. 
%Therefore, it implies that, as $w_{\rm de}$ gets smaller, $M_\nu$ would become larger to compensate the variation caused by $w_{\rm de}$ on power spectrum.

\begin{figure}[ht]
	\centering
	\includegraphics[width=0.5\textwidth]{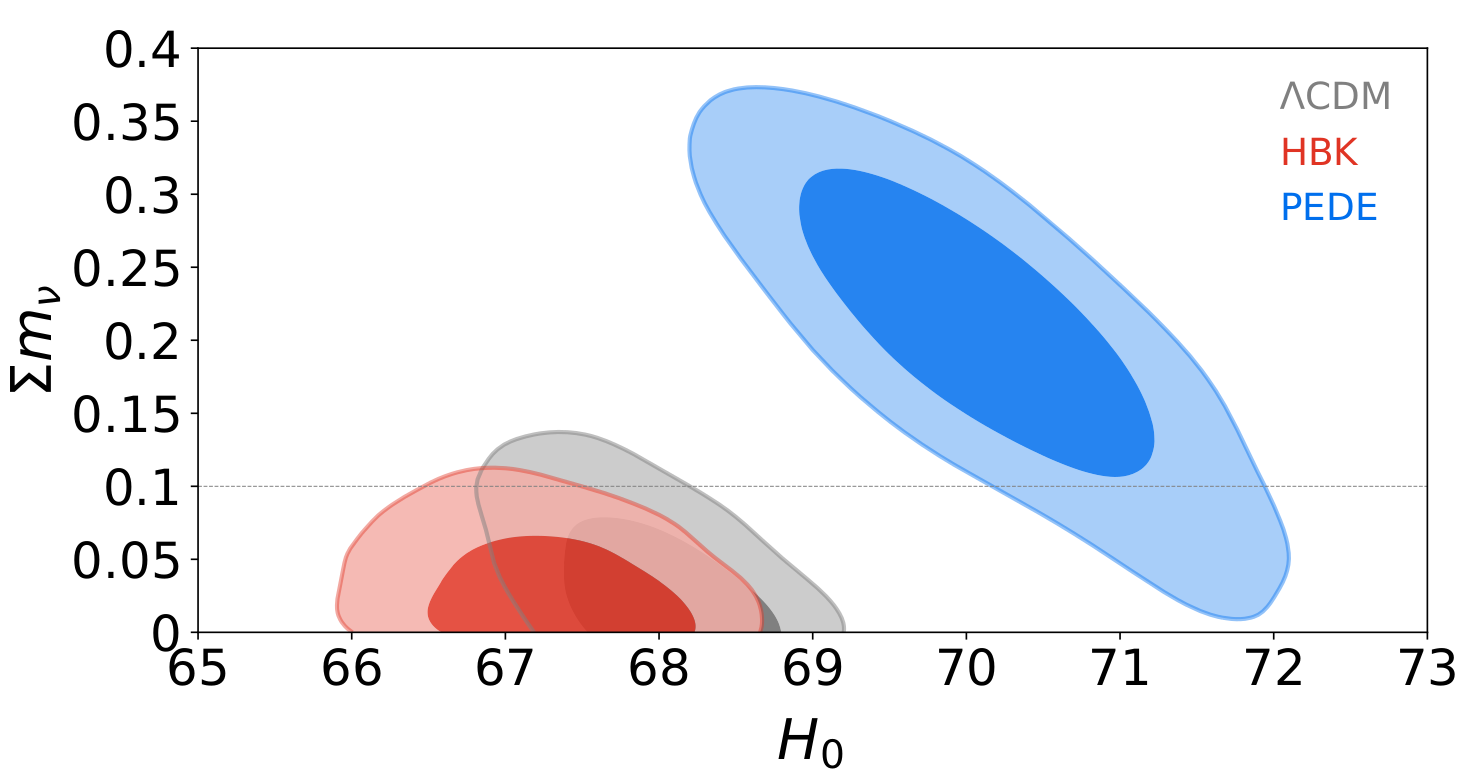}
	\caption{Marginalized 68.3\% CL and 95.4\% CL constraints on $H_0$ and $M_\nu$.}
	\label{hubblemnu}
\end{figure}

From only CMB data, there is a strong anti-correlation between Hubble constant $H_0$ and $M_\nu$, however, the degeneracy would be broken more effectively when combined with BAO and SNe data~\cite{2019arXiv190712598C}. In Fig.~\ref{hubblemnu}, we show the marginalized 68.3\% CL and 95.4\% CL constraints on $H_0$ and $M_\nu$. Compared with the results from Li et al. \cite{2019ApJ...883L...3L}, we can note that releasing $M_\nu$ does not help to relieve the $H_0$ tension between low-redshift measurement and high-redshift measurement. However, when considering massive neutrinos for PEDE model, the Hubble tension can still be alleviated.

\begin{figure}[ht]
	\centering
	\includegraphics[width=0.5\textwidth]{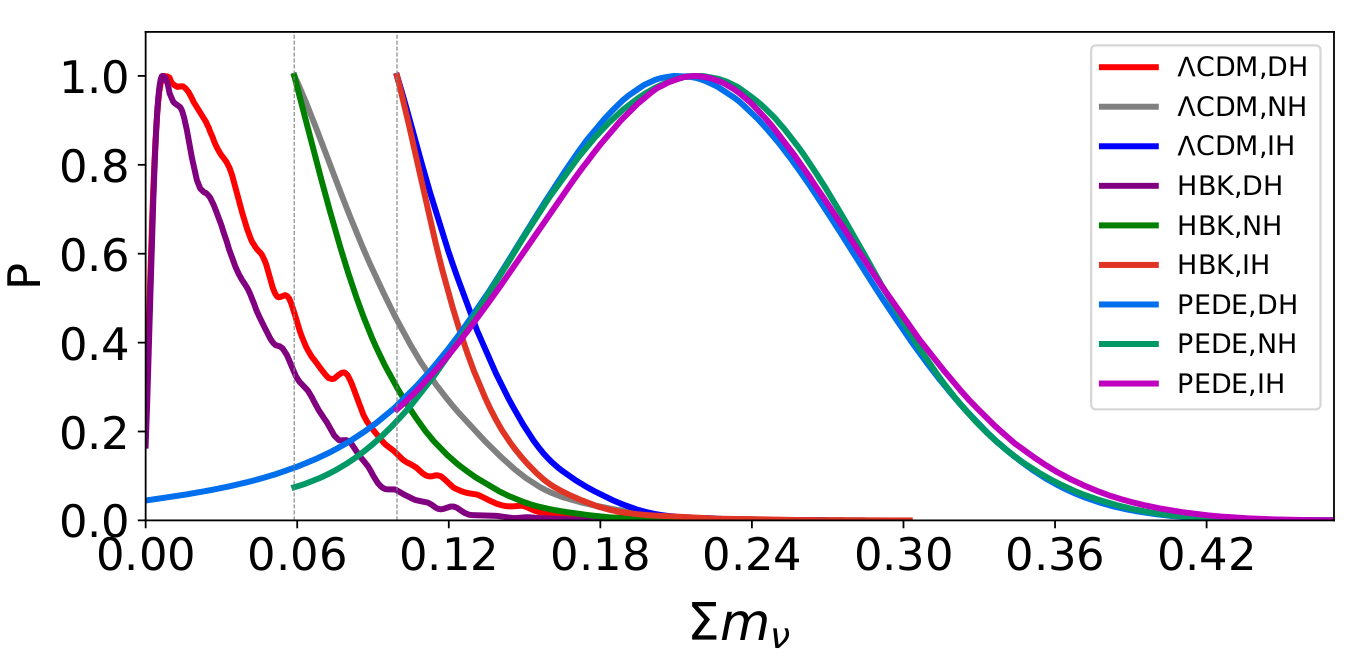}
	\caption{Comparison of 1-D marginalized posterior distribution for $M_\nu$ for three hierarchies within the three dark energy models.}
	\label{mnu}
\end{figure}

\subsection{Neutrino mass hierarchy}
In Tabs.~\ref{LCDMtab}, \ref{HBKtab} and \ref{PEDEtab}, we present the constraints of some selected parameters and the values of $\chi_{min}^2$ from MCMC analysis for $\Lambda$CDM, HBK and PEDE models. In Fig.~\ref{mnu}, we show comparison of 1-D marginalized posterior distribution on $M_\nu$ for three dark energy models with different hierarchies, which have been normalized. We can note that, though considering different hierarchies, the results that the quintessence prior of dark energy tends to tighten the cosmological constraint on $M_\nu$ still hold for each neutrino ordering. In PEDE model, the constraints on the total neutrino masses with different neutrino hierarchies are almost uniform. Within the datasets of CMB, BAO and SNe Ia, the PEDE model prefers higher values of $M_{\nu}$ compared to $\Lambda$CDM and HBK models, which leads to the masses of the individual neutrinos would be much larger than the squared mass differences. The larger individual neutrinos would blur the largest mass splitting influence. So it is difficult to distinguish different hierarchies for PEDE model.
%Whether the total mass is shared equally by the three neutrinos or it distributed according to one of the two hierarchies does not matter, since the individual masses are larger than the largest mass splitting.

\begin{table*}[ht]
%\begin{center}
\tbl{Marginalized constraints on cosmological parameters of the $\Lambda$CDM model for different neutrino hierarchies, which are given at 1$\sigma$ errors except the upper bounds on $M_\nu$ are given at 2$\sigma$ errors.}
{\begin{tabular}{c|c|c|p{3cm}}
\hline
\hline
 & Degenerate Hierarchy & Normal Hierarchy & Inverted Hierarchy\\
\hline
$\Omega_bh^2$  &$0.0224_{-0.0001}^{+0.0001}$&$0.0224_{-0.0001}^{+0.0001}$  & $0.0225_{-0.0001}^{+0.0001}$ \\
$\Omega_ch^2$   &$0.1193_{-0.0009}^{+0.0010}$ &$0.1191_{-0.0009}^{+0.0009}$ &$0.1189_{-0.0009}^{+0.0009}$ \\
$\Theta_s$    &$1.0140_{-0.0003}^{+0.0003}$&$1.0410_{-0.0003}^{+0.0003}$  &$1.0410_{-0.0003}^{+0.0003}$ \\
$\tau$   &$0.0554_{-0.0073}^{+0.0072}$ &$0.0574_{-0.0078}^{+0.0072}$ &$0.0592_{-0.0081}^{+0.0071}$ \\
$n_s$  &$0.9671_{-0.0038}^{+0.0037}$ &$0.9673_{-0.0037}^{+0.0037}$ &$0.9676_{-0.0036}^{+0.0037}$ \\
${\rm ln}(10^{10}A_s)$  &$3.0449_{-0.0143}^{+0.0143}$ &$3.0490_{-0.0151}^{+0.0150}$&$3.0527_{-0.0148}^{+0.0142}$ \\
$M_\nu$ (eV) &$\textless$ 0.114 &$\textless$ 0.150 & $\textless$ 0.170\\
$\sigma_8$  &$0.8145_{-0.0069}^{+0.0097}$ &$0.8044_{-0.0071}^{+0.0085}$ &$0.7982_{-0.0065}^{+0.0080}$ \\
$H_0$&$67.8936_{-0.4662}^{+0.4860}$&$67.5201_{-0.4590}^{+0.4565}$&$67.3298_{-0.4334}^{+0.4492}$\\
$\Omega_{m0}$&$0.3085_{-0.0061}^{+0.0060}$&$0.3127_{-0.0060}^{+0.0060}$&$_{-0.0060}^{+0.0058}$\\
$\chi_{min}^2$ &1910.795 &1911.990 &1913.325 \\
\hline
\end{tabular}}
%\end{center}
\label{LCDMtab}
\end{table*}

\begin{table*}[ht]
%\begin{center}
\tbl{Marginalized constraints on cosmological parameters of the HBK model for different neutrino hierarchies, which are given at 1$\sigma$ errors except the upper bounds on $M_\nu$ and the dark energy parameters are given at 2$\sigma$ errors.}
{\begin{tabular}{c|c|c|p{3cm}}
\hline
\hline
 & Degenerate Hierarchy & Normal Hierarchy & Inverted Hierarchy\\
\hline
$\Omega_bh^2$ &$0.0225_{-0.0001}^{+0.0001}$ &$0.225_{-0.0001}^{+0.0001}$  &$0.0225_{-0.0001}^{+0.0001}$ \\
$\Omega_ch^2$ &$0.1188_{-0.0010}^{+0.0010}$ &$0.1188_{-0.0009}^{+0.0009}$ &$0.1185_{-0.0009}^{+0.0009}$ \\
$\Theta_s$    &$1.0411_{-0.0003}^{+0.0003}$&$1.0411_{-0.0003}^{+0.0003}$  &$1.0411_{-0.0003}^{+0.0003}$ \\
$\tau$        &$0.0573_{-0.0080}^{+0.0071}$&$0.0588_{-0.0079}^{+0.0066}$ &$0.0607_{-0.0084}^{+0.0066}$ \\
$n_s$         &$0.9681_{-0.0039}^{+0.0037}$ &$0.9680_{-0.0034}^{+0.0033}$ &$0.9687_{-0.0037}^{+0.0037}$ \\
${\rm ln}(10^{10}A_s)$  &$3.0487_{0.0157}^{+0.0138}$ &$3.0514_{-0.0153}^{+0.0136}$&$3.0545_{-0.0162}^{+0.0134}$ \\
$M_\nu$ (eV)      &$\textless$ 0.087&$\textless$ 0.136&$\textless$ 0.164 \\
$\epsilon_s$  &$\textless$ 0.1902&$\textless$ 0.1727&$\textless$ 0.1896\\
$\epsilon_{\phi\infty}$  &$\textless$ 0.2988&$\textless$ 0.2732&$\textless$ 0.2371 \\
$\zeta_s$     &-&-&-\\
$\sigma_8$    &$0.8071_{-0.0082}^{+0.0094}$ &$0.7986_{-0.0076}^{+0.0093}$ &$0.7920_{-0.0075}^{+0.0083}$ \\
$H_0$         &$67.2687_{-0.5198}^{+0.5929}$&$67.0104_{-0.5083}^{+0.5498}$&$66.8172_{-0.5097}^{+0.5529}$\\
$\Omega_{m0}$  &$0.3130_{-0.0068}^{+0.0061}$&$0.3166_{-0.0062}^{+0.0062}$&$0.3189_{-0.0063}^{+0.0064}$\\
$\chi_{min}^2$   &1910.252 &1911.919 & 1913.514\\
\hline
\end{tabular}}
%\end{center}
\label{HBKtab}
\end{table*}

\begin{table*}[ht]
%\begin{center}
\tbl{Marginalized constraints on cosmological parameters of the PEDE model for different neutrino hierarchies, which are given at 1$\sigma$ errors and also given at 2$\sigma$ errors on $M_\nu$.}
{\begin{tabular}{c|c|c|p{3cm}}
\hline
\hline
 & Degenerate Hierarchy & Normal Hierarchy & Inverted Hierarchy\\
\hline
$\Omega_bh^2$  &$0.0223_{-0.0001}^{+0.0001}$ &$0.0223_{-0.0001}^{+0.0001}$  &$0.0223_{-0.0001}^{+0.0001}$ \\
$\Omega_ch^2$   &$0.1212_{-0.0009}^{+0.0009}$ &$0.1212_{-0.0010}^{+0.0009}$ &$0.1212_{-0.0009}^{+0.0009}$ \\
$\Theta_s$    &$1.0407_{-0.0003}^{+0.0003}$&$1.0407_{-0.0003}^{+0.0003}$  & $1.0407_{-0.0003}^{+0.0003}$\\
$\tau$   &$0.0530_{-0.0075}^{+0.0072}$ &$0.0535_{-0.0075}^{+0.0068}$ & $0.0541_{-0.0074}^{+0.0072}$ \\
$n_s$  &$0.9619_{-0.0038}^{+0.0036}$ &$0.9622_{-0.0036}^{+0.0036}$ & $0.9619_{-0.0036}^{+0.0037}$\\
${\rm ln}(10^{10}A_s)$  &$3.0444_{-0.0147}^{+0.0148}$ &$3.0453_{-0.0147}^{+0.0134}$&$3.0464_{-0.0153}^{+0.0151}$ \\
$M_\nu$ (eV)  &$0.2123_{-0.0642-0.1367}^{+0.0666+0.1293}$ &$0.2188_{-0.0648-0.1257}^{+0.0646+0.1238}$ & $0.2257_{-0.0711-0.1249}^{+0.0574+0.1094}$ \\
$\sigma_8$  &$0.8239_{-0.0164}^{+0.0166}$ &$0.8227_{-0.0162}^{+0.0164}$ &$0.8213_{-0.0145}^{+0.0174}$ \\
$H_0$&$70.0558_{-0.7679}^{0.7285}$&$69.9905_{-0.7098}^{+0.7851}$&$69.9210_{-0.7230}^{+0.7052}$\\
$\Omega_{m0}$&$0.2971_{-0.0081}^{+0.0079}$&$0.2977_{-0.0087}^{+0.0074}$&$0.2985_{-0.0076}^{+0.0077}$\\
$\chi_{min}^2$ & 1918.044 &1918.425 & 1917.067\\
\hline
\end{tabular}}
%\end{center}
\label{PEDEtab}
\end{table*}

\begin{figure*}[ht]
    \centering
    \includegraphics[width=0.8\textwidth]{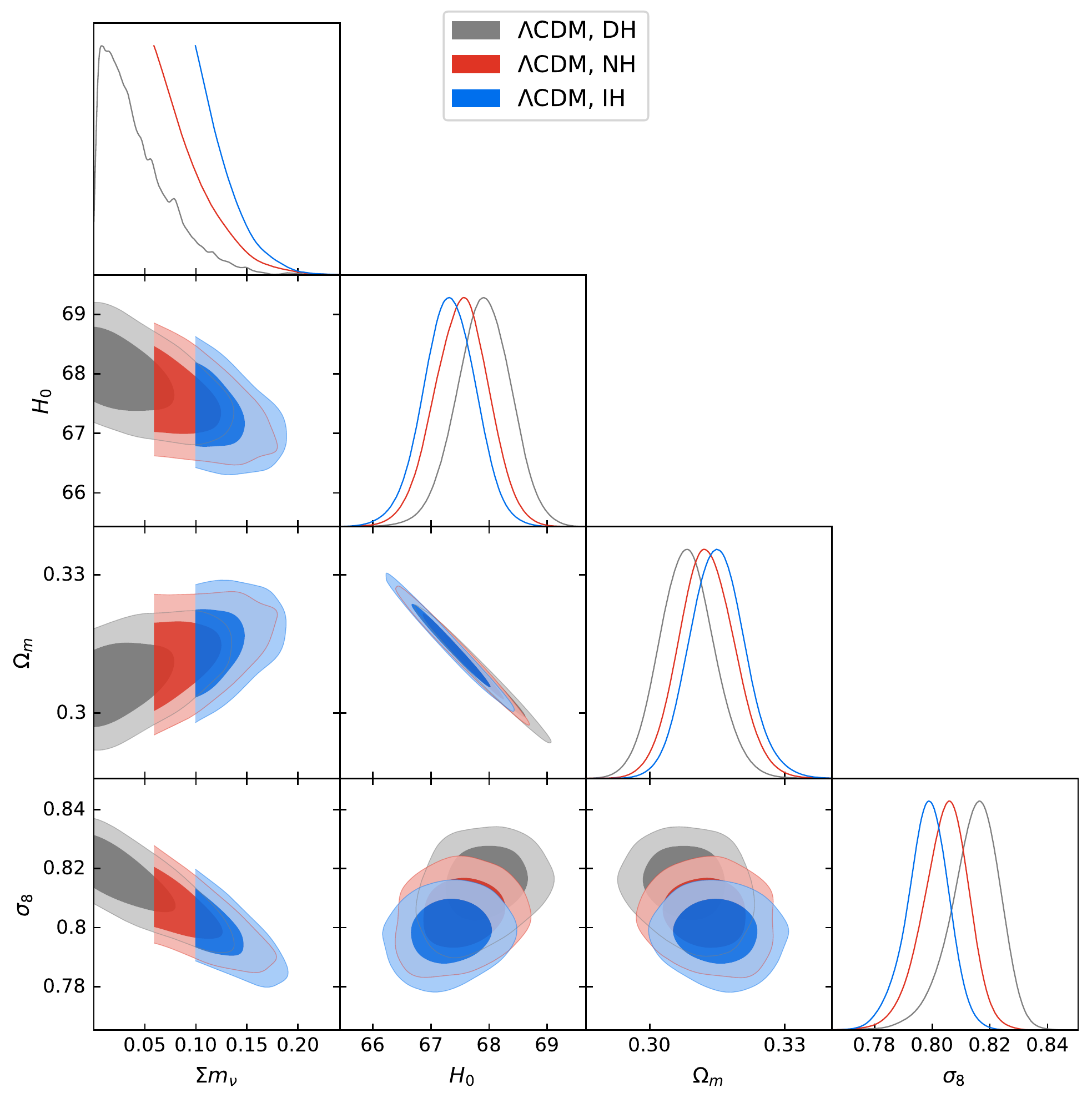}
    \caption{68\% and 95\% CL contour plot in the $\Lambda$CDM model.}
    \label{LCDMfig}
\end{figure*}

\begin{figure*}[ht]
    \centering
    \includegraphics[width=0.8\textwidth]{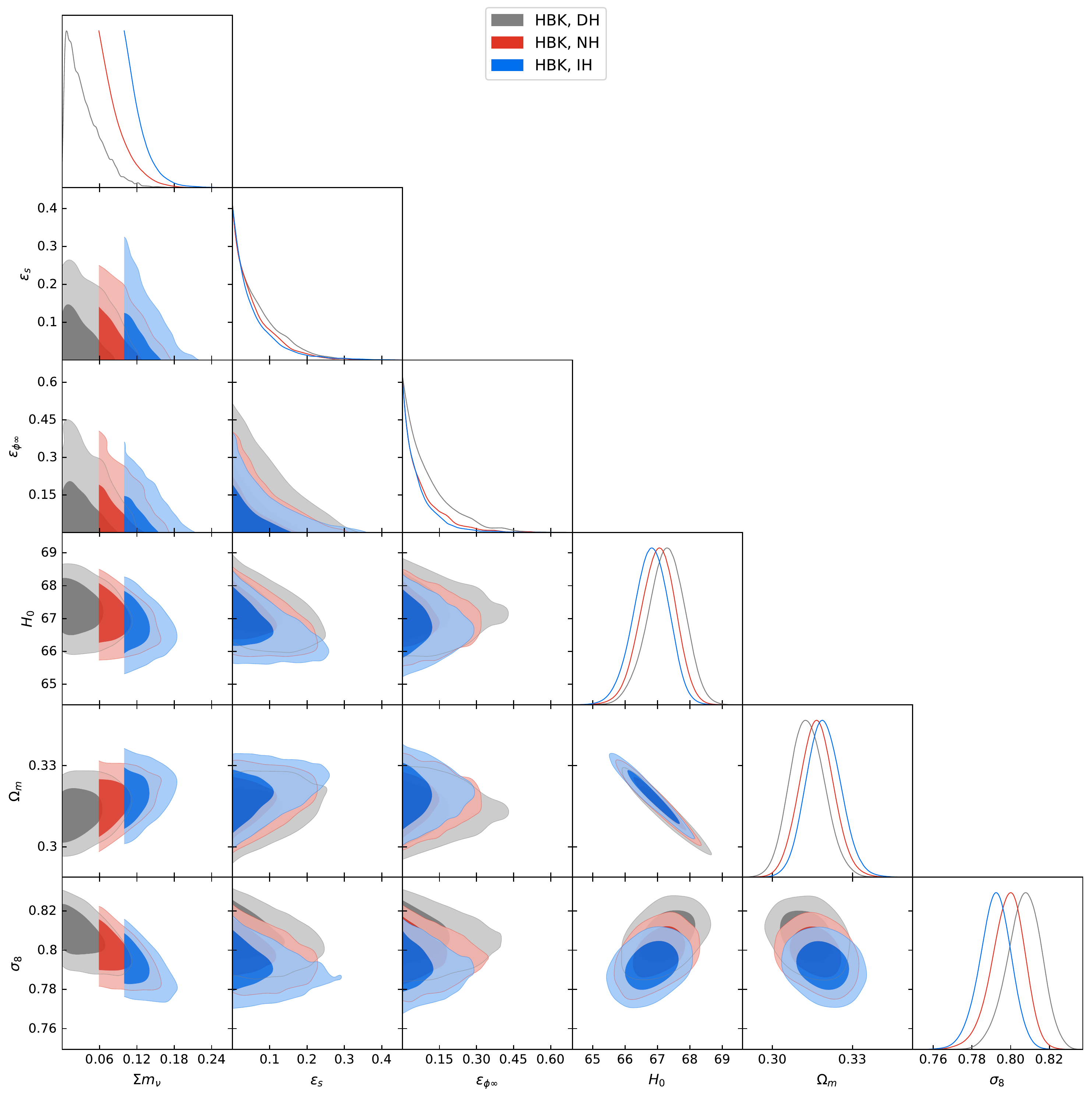}
    \caption{68\% and 95\% CL contour plot in the HBK model.}
    \label{HBKfig}
\end{figure*}

\begin{figure*}[ht]
    \centering
    \includegraphics[width=0.8\textwidth]{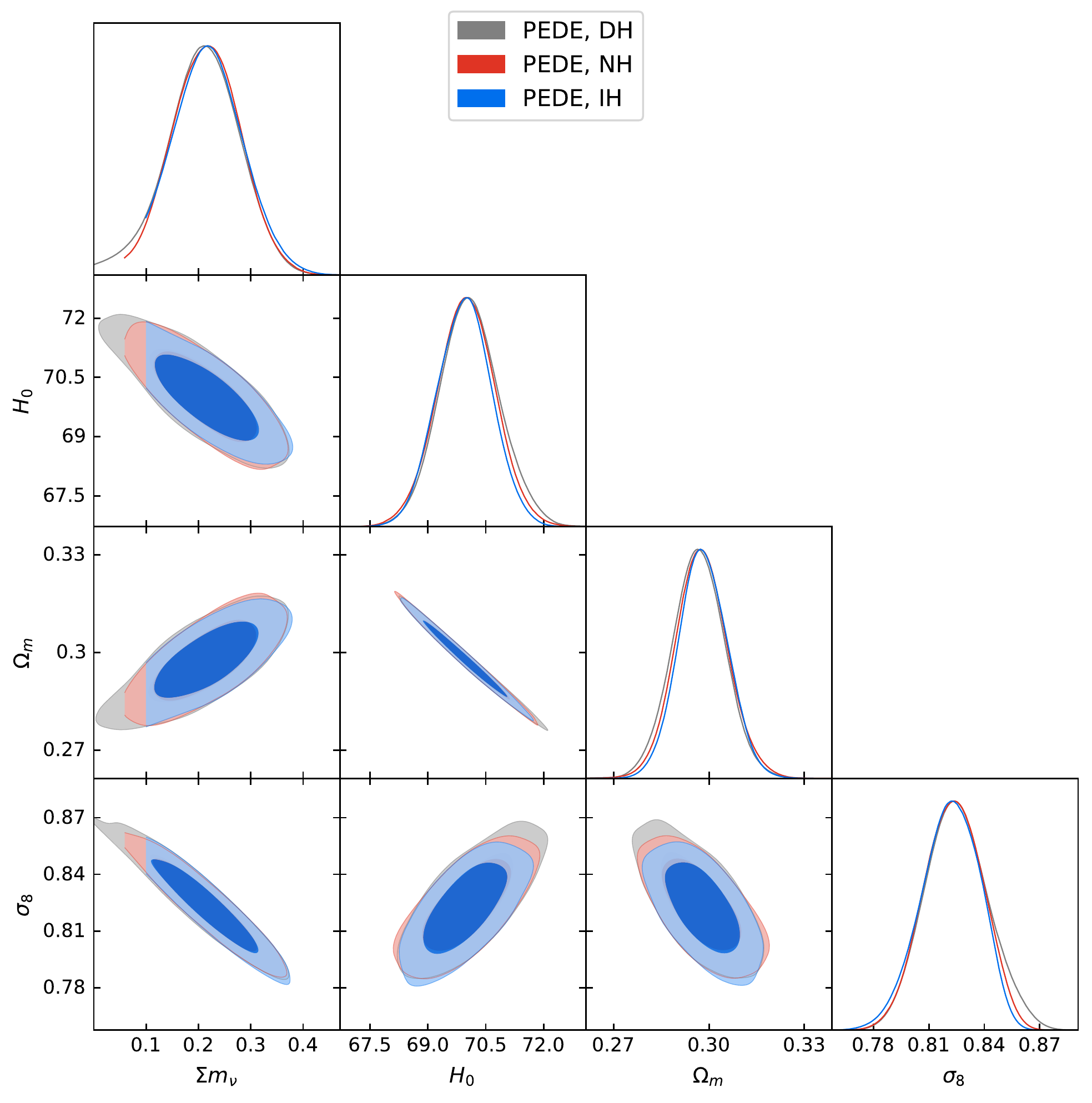}
    \caption{68\% and 95\% CL contour plot in the PEDE model.}
    \label{PEDEfig}
\end{figure*}

Figs. \ref{LCDMfig}, \ref{HBKfig} and \ref{PEDEfig} depict the 1D marginalized posterior distributions and 2D joint contours at 68\% and 95\% CL for some selected cosmological parameters of the $\Lambda$CDM, HBK and PEDE models. In the $\Lambda$CDM and HBK models, different hierarchies lead to some slight effects on other cosmological parameters because of the degeneracy, as shown in Fig.~\ref{LCDMfig} and Fig.~\ref{HBKfig}. For instance, as the value of $M_\nu$ increases from the degenerate approximation to normal hierarchy to inverted case, the value of $H_0$ and $\sigma_8$ decrease and $\Omega_m$ increase. However, these variations are not significant enough and the impact of neutrino hierarchy on dark energy parameters is only at a few percent level. In PEDE model, because the constraints on $M_\nu$ are similar in three hierarchies, all the cosmological parameters almost coincide. Therefore, with the improvement of observation accuracy, we may not be able to solve the neutrino mass hierarchy problem for PEDE model in the future. Besides, we present the value of $\chi_{min}^2$ calculated at best-fit points for each case. Current cosmological observations can not provide a rigorous statistical treatment for hierarchy preference. The differences between the values of $\chi_{min}^2$ for normal hierarchy and for inverted hierarchy are not significant. However, there is a kind of interesting point that the PEDE model slightly prefers IH, which is different from the $\Lambda$CDM and HBK models which slightly prefer NH. Besides, we notice that the $\chi^2_{min}$ in PEDE model is larger than the $\chi^2_{min}$ in $\Lambda$CDM. The $\Lambda$CDM and the PEDE model have the same parameters. From Fig. \ref{w_de}, we know the $w_{de} = -1$ for $\Lambda$CDM and it deviates a lot from $-1$ for PEDE model. The datasets from CMB, BAO and SNe Ia are more likely to have a $w_{de}$ closer to $-1$.

\begin{figure*}[!htb]
    \centering
    \includegraphics[width=0.8\textwidth]{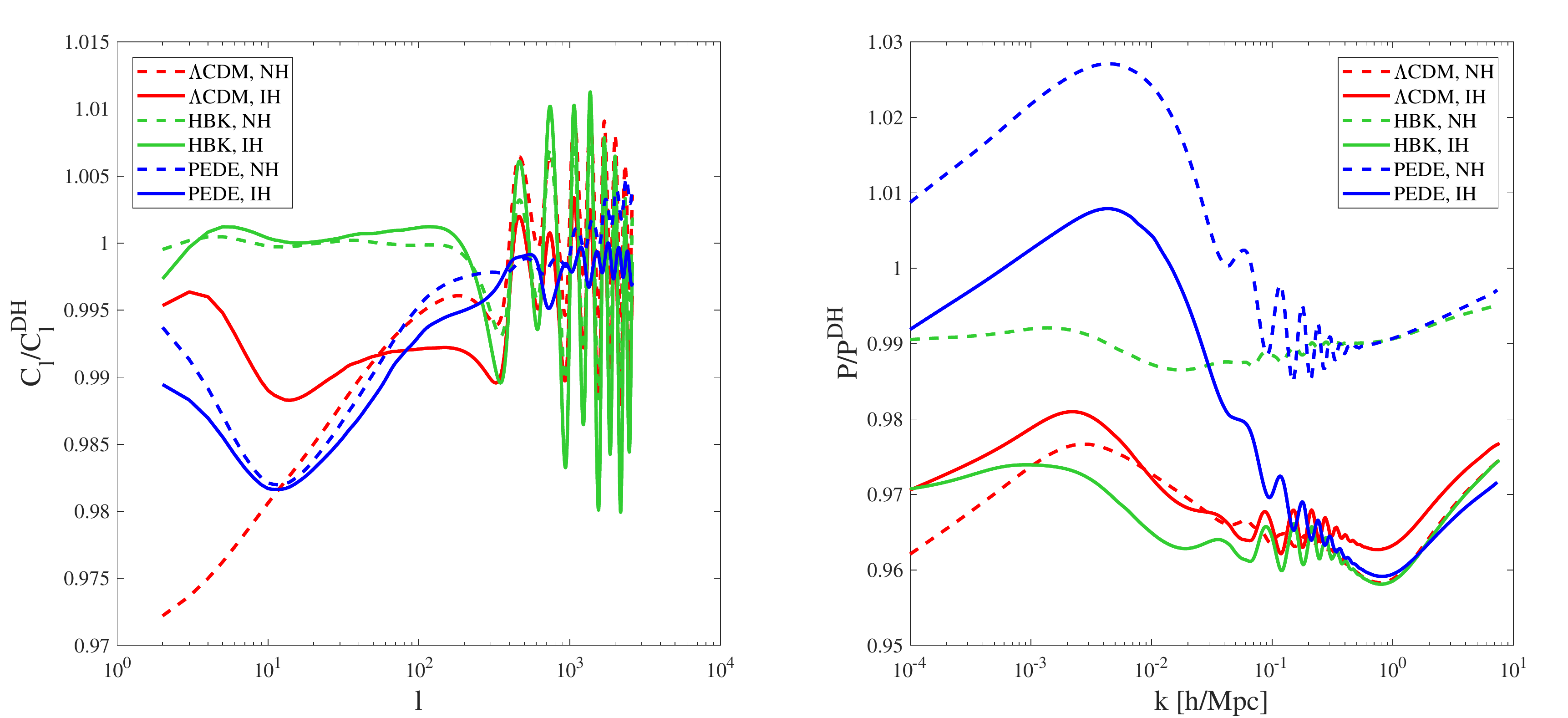}
     \caption{The ratio of NH and IH to DH of the CMB temperature power spectrum $C_l^{TT}$ and  the matter power spectrum for PEDE, $\Lambda$CDM and HBK models.}
    \label{allCls}
\end{figure*}

We also show the theoretical predictions on CMB temperature spectrum and matter power spectrum for the three models with three hierarchies in Fig. \ref{allCls}. We assumed the best-fit values generated by MCMC analysis above. When the neutrino hierarchies are considered, the ratio of the spectrum with NH or IH to the spectrum with DH is very close to 1, the impact of neutrino hierarchy on the power spectrum is only at a few percent level. Therefore, neutrino hierarchy has very slight impacts on other cosmological parameters whichever type of dark energy we apply.

\section{Conclusion}\label{four}

Cosmology can be used to address the problems about the total neutrino masses and the mass hierarchy. Current observations can provide constraints on the sum of the neutrino masses $M_\nu$. However, the bounds on $M_\nu$ from cosmology are model-dependent. In this work, we investigated cosmological constraints on $M_\nu$ within the $\Lambda$CDM model, the HBK model and PEDE model. These models are well constrained by the latest observational data, CMB + BAO + SNe Ia. For $\Lambda$CDM model, when we use the latest Planck 2018 CMB data, we obtain tighter upper bounds on $M_\nu$. For HBK model, we find the quintessence prior of dark energy tends to tighten the cosmological constraints on $M_\nu$, as previously stated. On the other hand, the phantom prior of PEDE model tends to make the constraints on $M_\nu$ looser and its value larger, and we also obtain a nonzero lower bounds (95\%CL) on $M_\nu$.

In addition, we also consider the impacts of different neutrino hierarchies for the three models. It leads to some effects on cosmological parameters due to the variations of $M_\nu$. However, the changes to any of these parameters are not significant enough. Especially, in the PEDE model, the change of neutrino hierarchy nearly has no impacts on other cosmological parameters.

\section*{Acknowledgments}
First, we would like to express gratitude to our tutor, Prof. Zhiqi Huang.  He usually provided us valuable advice and enlightenment for ideas. Additionally, we would like to thank Miaoxin Liu for providing us technical advisory assistance.

%%%%%%%%%%%%%%%%%%%%%%%%%%%%%%%
%\vfill
\bibliographystyle{ws-ijmpd.bst}
\bibliography{cites.bib}
%%%%%%%%%%%%%%%%%%%%%%%%%%%%%%%

\end{document}